\documentclass[preprint,review,12pt,times,authoryear]{elsarticle}

\usepackage[english]{babel}
\usepackage[utf8]{inputenc}
\usepackage{graphicx}
\usepackage{amssymb}
\usepackage{amsmath}
\usepackage{subfigure}

\usepackage{lineno}
\usepackage{setspace}

\usepackage{color}
\newcommand{\revised}{\color{blue}}

\journal{M\&M}

\begin{document}

\begin{frontmatter}

\title{Reflections on the projection of ions in atom probe tomography}

\author[simap1,simap2]{Frédéric De Geuser}
\ead{frederic.de-geuser@simap.grenoble-inp.fr}
\author[mpie]{Baptiste Gault}
\ead{b.gault@mpie.de}

\address[simap1]{Univ. Grenoble Alpes, SIMAP, F-38000 Grenoble, France}
\address[simap2]{CNRS, SIMAP, F-38000 Grenoble, France}
\address[mpie]{Max-Planck Institut für Eisenforschung, Max-Planck-Straße 1,
D-40237 Düsseldorf, Germany}

\begin{abstract}
There are two main projections used to transform, and reconstruct, field ion micrographs or atom probe tomography data into atomic coordinates at the specimen surface and, subsequently, in three-dimensions. In this article, we present a perspective on the strength of the azimuthal equidistant projection in comparison to the more widely used and well-established point-projection {\revised (or pseudo-stereographic projection)}, which underpins data reconstruction in {\revised most software packages currently in use across the community}. After an overview of the reconstruction methodology, we demonstrate that the azimuthal equidistant is {\revised more robust } with regards to errors on the parameters used to perform the reconstruction and is therefore more likely to yield more accurate tomographic reconstructions.
\end{abstract}


\end{frontmatter}

\section{Introduction}
Since the introduction of the field-emission electron microscope (FEEM) and subsequently of the field-ion microscope (FIM), it has been recognized that they are projection microscopes that provide a highly-magnified image of the surface of the specimen. {\revised This projection also underpins atom-probe tomography (APT) and enables the technique to analyze individual features at the sub-nanometer scale.}

The high {\revised spatial} resolution of the FIM and its capacity to resolve individual atomic terraces has made it a tool of choice for investigating the true nature of the projection. A field ion micrograph is indeed a projected image of the specimen surface, revealing details of its structure down to the atomic level. In the 1960s and 1970s,  \citet{Brandon1964c} and \citet{Newman1967}, among others, reported on the study of the properties of the projection in field ion microscopy. They compared micrographs to known point-projections, such as gnomonic or stereographic, only to show that none of those projections often used in other microscopy techniques could be used as such. Based on the arguments of Newman, \citet{Wilkes1974} proposed that the projection led to features separated by an angle $\theta$ at the specimen surface would be separated by a distance $\rho=k_{\theta} \times \theta$ at the detector. This was subsequently referred to as a \emph{linear projection}. 

From the 1980's, the community's focus progressively shifted to atom-probe microanalysis and then to atom probe tomography. The progressive adoption of the reconstruction protocol, proposed by Blavette, Bostel and co-workers \citep{Blavette1982,bas_general_1995} and subsequently modified to remove small-angle approximations \citep{Geiser2009a,gault_advances_2011}, across the entire atom probe user base has led to a premature end of the discussions on this topic. However, as the performance of APT are routinely challenged, particularly in the analysis of complex materials, it is important to revisit the properties of the ion projection, and its intimate link to the specimen geometry. 

In 1999, Cerezo and co-workers revisited the \emph{linear projection} \citep{Cerezo1999a}, and highlighted once again that it was a better representation of the actual ion projection in FIM than a pseudo-stereographic projection, but also tested its limits. Two interesting points arose from this study: the \emph{linear projection} holds through a tilt series, or if the specimen is rotated about its main axis; however, the proportionality factor between the angle at the specimen surface and the distance on the detector cannot be assumed constant across all possible combinations of poles. 

Cerezo's work pointed to the idea that a center of the projection must be defined and that the azimuths are maintained through the projection: the ions fly within a plane that contains the main specimen axis and its original position at the specimen surface. If we assume azimuths are conserved, then we are dealing with azimuthal equidistant projection, which is a well-known and well-described projection used, e.g. by geographers \citep{Snyder1997}, and is the projection model of the Earth represented on the United Nations emblem. {\revised More recently, \citet{Miller2014} also discussed the Hawkes-Kasper approach, which is an equivalent to this projection model.}

In this article, we provide a critical {\revised viewpoint of the main projection models}, showcase how a precise determination of the orientation of the specimen can be achieved via the identification of the crystallographic features present in the detector hit map, and finally introduce a framework that could be generalized and form the base for a new reconstruction paradigm, which is described and discussed.

\section{General reconstruction framework}
\subsection{System of coordinates}
The reconstruction protocol in APT consists in a transformation of the detector coordinates $(X_D,Y_D)$ and detection sequence $N$, as well as an instantaneous radius of curvature $R$ into the coordinates of individual atoms $(x,y,z)$ in the tomographic reconstruction built assuming that the specimen is a spherical cap sitting on a truncated cone \citep{bas_general_1995,gault_advances_2011}:
\begin{equation}
(X_D,Y_D,N,R)\Longleftrightarrow (x,y,z)
\end{equation}

\begin{figure*}[htbp]
	\centering
		\includegraphics[width=8cm]{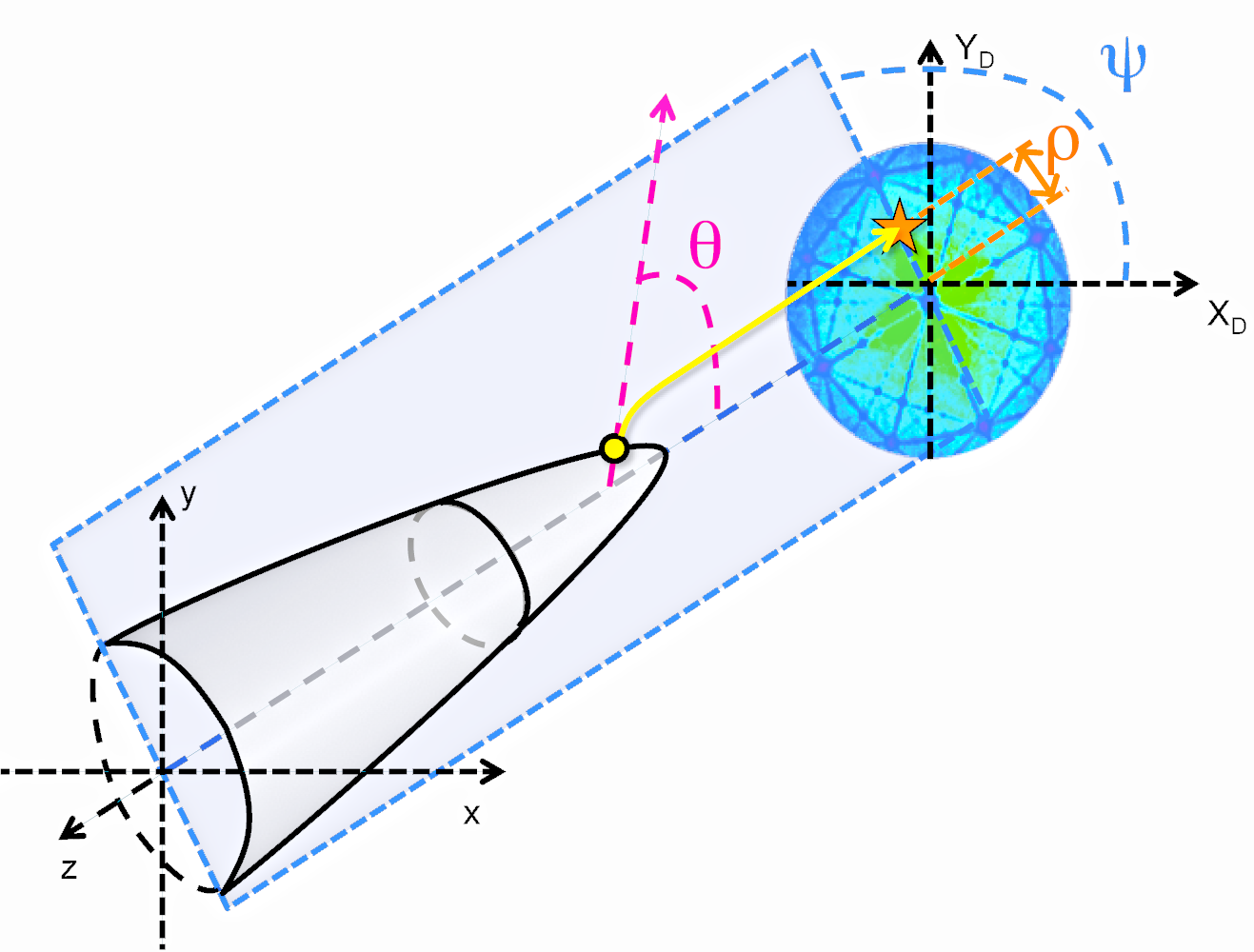}
	\caption{\label{schematic}Schematic view of the emitter and system of coordinates used in this study (not to scale).}
\end{figure*}

A point that was introduced by \citet{gault_advances_2011} is the use of cylindrical coordinates to describe the  system. We can write the relationship between cylindrical and Cartesian coordinates (Fig.~\ref{schematic}) as:

\begin{equation}
\begin{cases}
X_D=\rho\cos\psi\\
Y_D=\rho\sin\psi
\end{cases}
\label{coordDetec}
\end{equation}
on the detector, and
\begin{equation}
\begin{cases}
x=r\cos\psi=R\sin\theta\cos\psi\\
y=r\sin\psi=R\sin\theta\sin\psi\\
z=z_c(N)+R\cos\theta
\end{cases}
\label{coordSample}
\end{equation}
at the specimen surface.

$R$ is the radius of the specimen, $\theta$ is the launch angle, $\psi$ is the azimuthal angle, and $z_c$ is the $z$ coordinate of the centre of the apex of the spherical cap, which can conventionally be set to 0 at the beginning of the experiment, but varies with the sequence of evaporation. $z_c$ is a function of the detection sequence and can be seen as the analyzed depth.

In this framework, an ion strikes the detector at the coordinate $(\rho,\psi)$ originating from a position at the surface of the specimen characterized by $(r,\psi,z)$ or $(R,\theta,\psi,z_c)$.  For the remaining of this study, we will assume that the projections of the angle is \emph{azimuthal}. This implies that there exist a particular point on the detector plane (but possibly outside the physical detector) around which the azimuthal angles (e.g. $\psi$) are preserved. This point will be called the \emph{center of the projection} or \emph{projection center}. The pseudo-stereographic projection model, for instance, is an azimuthal projection, the center of which is at the intersection of the axis of the specimen and the detector plane. Within this assumption, we can choose this point as the origin of the $(\rho,\psi)$ detector coordinate so that $\psi$ is preserved (which was implied in the use of the same symbol in eq.~\ref{coordDetec} and \ref{coordSample}).

\subsection{Tomographic reconstruction}
{\revised The transformation of the detector coordinates $(\rho,\psi,N)$ into the $(\theta,\psi,z_c)$ makes then use of the radius of curvature $R$}. $\psi$ is unchanged through an azimuthal projection so that we are left with
\begin{enumerate}
\item  computing $\theta$ from $\rho$, i.e. adopting an \emph{angular projection} model
\item  computing $R$, i.e. adopting an \emph{radius evolution} model
\item computing $z_c$ from $N$, $\theta$ and $R$, i.e. adopting a \emph{depth of analysis} model
\item computing  $(x,y,z)$ through equation~(\ref{coordSample})
\end{enumerate}
 
This paper is concerned with the first item, namely the angular projection model. We will thus assume that we use a satisfactory radius evolution model as well as a robust model for the depth increment. 
 
{\revised Classical methods to model evolution of the specimen radius are based either on the usual relationship $R=V/(k_fF_{ev})$ (usually referred to as \emph{voltage reconstruction mode}), on the change in radius constrained by a constant shank angle, or on a micrograph of the specimen profile (respectively \emph{shank angle} or \emph{tip profile reconstruction mode} in the most commonly used commercial data treatment software)}. In any case, one or more additional parameters must be introduced: $k_f$ for voltage mode, initial radius ($R_0$) and the half-shank angle ($\alpha$) for the \emph{shank angle} mode or a complete profile of the specimen derived from a high-resolution electron micrograph of its outer shape.

Finally, the depth of analysis $z_c$ of the current ion $N$ can be deduced from that of the previous ion $N-1$ through the following relationship:
\begin{equation}
z_c\left(N\right)=z_c\left(N-1\right)+\frac{\mathrm{d}z_c}{\mathrm{d}N}
\end{equation}
$\frac{\mathrm{d}z_c}{\mathrm{d}N}$ is the so-called depth increment, which was defined by \citep{bas_general_1995} as the (virtual) depth of a layer corresponding to the removal of a single atom. Its value is determined in such a way that the total analyzed volume is necessarily equal to the total number of atom multiplied by the atomic volume (corrected for the limited detection efficiency $\eta$). A way\footnote{non unique, but always used} to achieve this is to attribute to each atom a volume increment corresponding to the atomic volume $\Omega$:
\begin{equation}
\frac{\mathrm{d}V}{\mathrm{d}N}=\frac{\mathrm{d}V}{\mathrm{d}z_c}\frac{\mathrm{d}z_c}{\mathrm{d}N}=\frac{\Omega}{\eta}
\end{equation}
so that
\begin{equation}
\frac{\mathrm{d}z_c}{\mathrm{d}N}=\frac{\Omega}{\eta\frac{\mathrm{d}V}{\mathrm{d}z_c}}
\end{equation}
$\frac{\mathrm{d}V}{\mathrm{d}z_c}$ has the units of a surface and can be seen as the projected analyzed surface. If the effective detected surface on the detector is a disk and the maximum observable angle is $\theta_{max}$, then $\frac{\mathrm{d}V}{\mathrm{d}z_c}=\pi R^2\sin^2\theta_{max}$. If the effective detection surface is not exactly a disk, which is often the case when the end-user selects a specific area of the detector for data reconstruction, a more general formula would be needed that includes a geometrical factor $\sigma$ depending only on the shape of the selected area:
\begin{equation}
\frac{\mathrm{d}z_c}{\mathrm{d}N}=\frac{\Omega}{\eta \sigma \pi \sin^2\theta_{max} R^2}\label{increment}
\end{equation}

Interestingly, $\theta_{max}$ depends on the angular projection model, which is itself strongly dependent on the specimen and microscope geometries. Therefore, $\theta_{max}$ is not a constant, and actually is expected to evolve throughout a single analysis. Finally, please note that the formula introduced in \citet{gault_advances_2011} to extend the concept introduced by \citet{bas_general_1995} equating $\frac{\mathrm{d}V}{\mathrm{d}z_c}$ to the reverse projection of the detector onto the surface of the spherical cap is not generally applicable. 

\section{Angular projection models}
If the specimen is assumed to possess a spherical end-shape, the position of an atom at the surface of the sample can be given by its \emph{longitude} and \emph{latitude} angles, widely used in geographical mapping. In the notation of this work, the longitude is $\psi$. Since we consider only azimuthal projections, its value is conserved through the projection. While the latitude represents the elevation from the equatorial plane, in atom probe tomography, it is more conventional to access it through its complementary angle $\theta$ which gives the angle between the axis of the specimen and the position of the atom.

Defining an angular projection is simply finding the relationship between $\rho$, the distance between the projection center within the detector plane and the detected hit position, and $\theta$, the launch angle, i.e. $\rho=f(\theta)$. 

\subsection{Pseudo-stereographic projections (standard model)}

The pseudo-stereographic projection is the \emph{de facto} standard angular projection model for atom probe tomography. It is a point-projection model. It can be envisioned as an intermediate situation between a \emph{gnomonic} projection (i.e. where the origin is the center of the spherical cap) and a \emph{stereographic projection} (i.e. where the origin is at the south pole, i.e. at twice the radius from the surface). In the pseudo-stereographic model, all trajectories originate from the same point situated on the specimen axis but behind the center of the spherical cap. The distance from the centre to the apparent origin of the trajectories is written $mR$. Trigonometric considerations leads to the following expression for the projection:
\begin{equation}
\rho=\frac{L}{m+\cos\theta}\sin\theta
\end{equation}
A possible misconception originating from the ubiquitous use of the expression \emph{image compression factor} is that the standard pseudo-stereographic model for atom probe reconstruction is equivalent to a linear compression of the launch angles so that the ions are detected with an apparent angle $\theta'$ such that
\begin{equation}
\theta=\xi\theta'
\end{equation}
with $\xi$ being the image compression factor (ICF).  In the pseudo-stereographic projection model, however,  $\xi=\theta/\theta'$ is not a constant over the detector. It can be expressed as:
\begin{equation}
\xi=\frac{\theta}{\mathrm{atan}\left(\frac{\sin\theta}{m+\cos\theta}\right)}
\end{equation}
While this leads to the expected $\xi=m+1$ for small $\theta$ and is still a reasonable assumption up to any practical angles, it is not strictly equivalent to a simple angular compression.

The inverse projection gives:
\begin{equation}
 \theta=\mathrm{atan}\left(\frac{\rho}{L}\right)+\mathrm{asin}\left(m\sin\left(\mathrm{atan}\left(\frac{\rho}{L}\right)\right)\right)
\end{equation}

\subsection{Azimuthal equidistant projections (linear model)}
As shown independently by \citet{Newman1967}, \citet{Wilkes1974} and \citet{Cerezo1999a} using FIM, the relationship between distance on the detector and crystallographic angle is in fact better reproduced by a linear relationship, i.e.
\begin{equation}
\rho=k\theta
\end{equation}
The convergence of the projections at small angles imposes that:
\begin{equation}
k=\frac{L}{\xi}=\frac{L}{m+1}
\end{equation}
but the projection models are only equivalent at small angles. The inverse projection is:
\begin{equation}
\theta=\frac{\rho}{k}
\end{equation}

In geographical mapping, this model is called azimuthal equidistant. Similarly to the pseudo-stereographic projection, it is azimuthal. It is equidistant in the sense that the distance on the detector $\rho$ (i.e. on the map) between the center and any other point is proportional to the actual distance at the surface of the specimen's spherical cap  (i.e. $R\theta$). The azimuthal equidistant projection is not a point projection and the apparent trajectories of the ions do not cross at a single projection point, as pointed out in the description of the ion projection in FIM by i.e. \citet{Newman1967}.

\subsection{Comparison of the models}

\subsubsection{Distance vs. angle}
\begin{figure}[htbp]
	\centering
		\subfigure{\includegraphics[height=6.5cm]{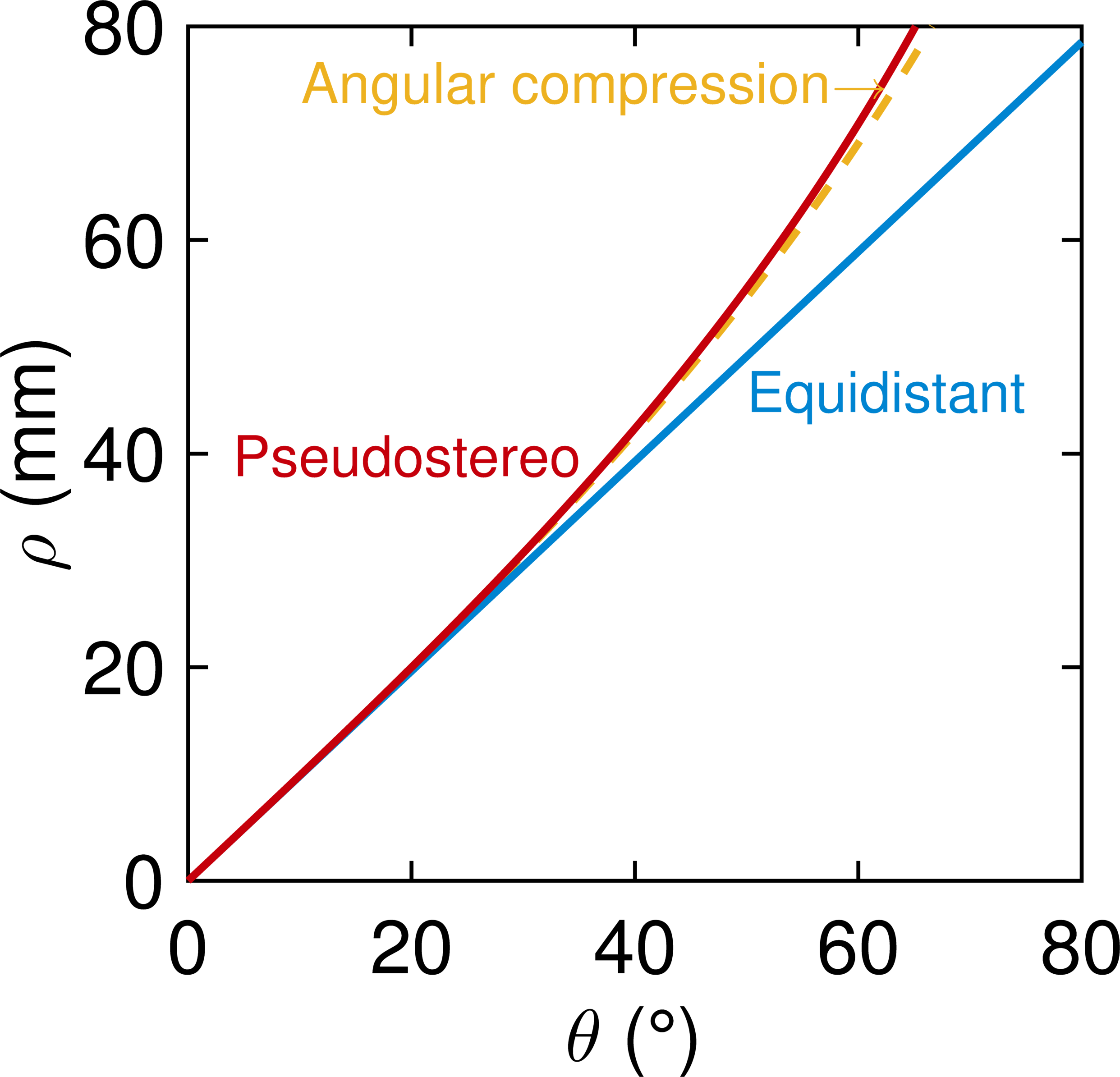}\label{comp_Dist_vs_angle}}
		
		\subfigure{\includegraphics[height=6.5cm]{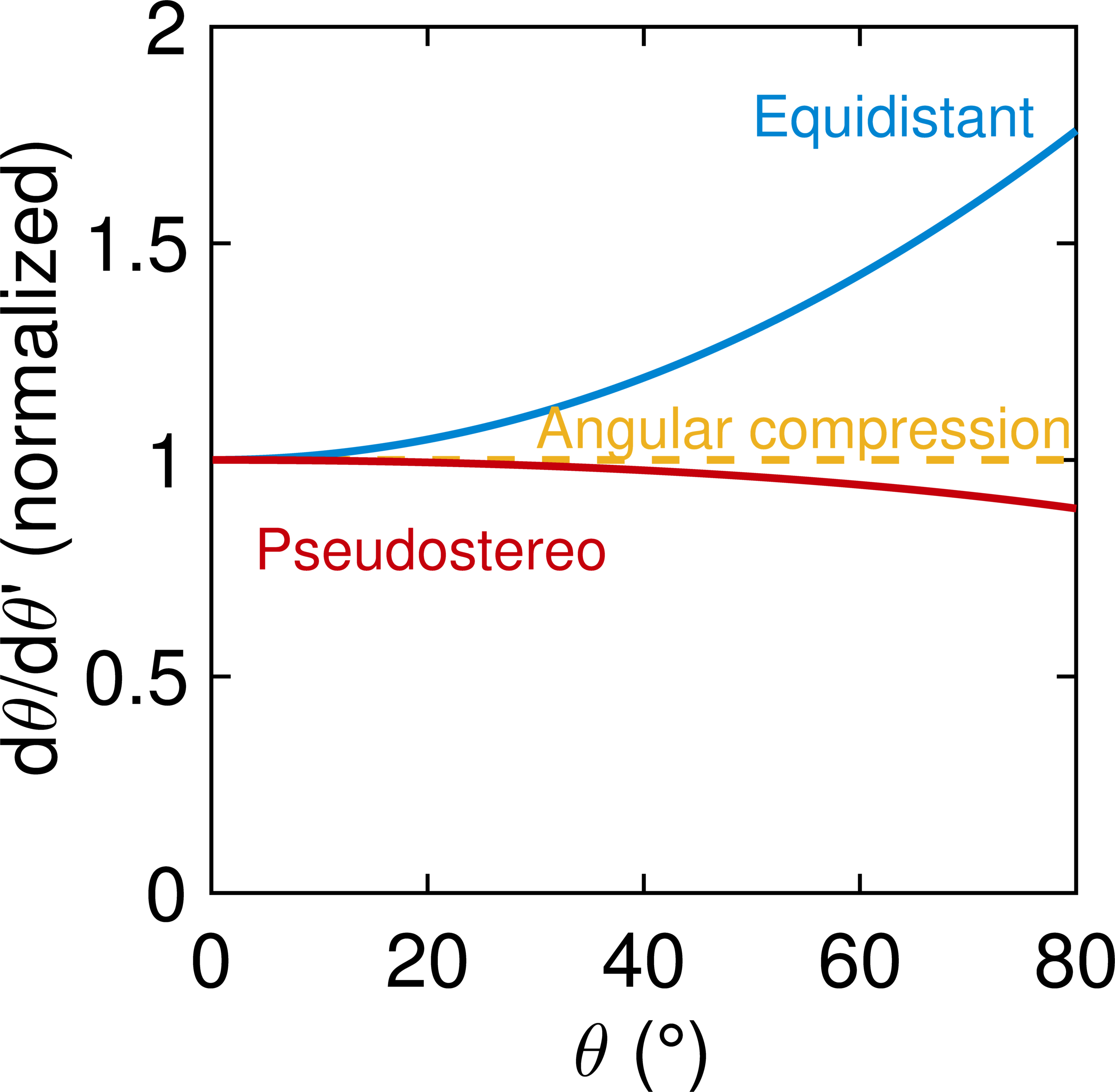}\label{comp_ICF}}
		\hfill		
		\subfigure{\includegraphics[height=6.5cm]{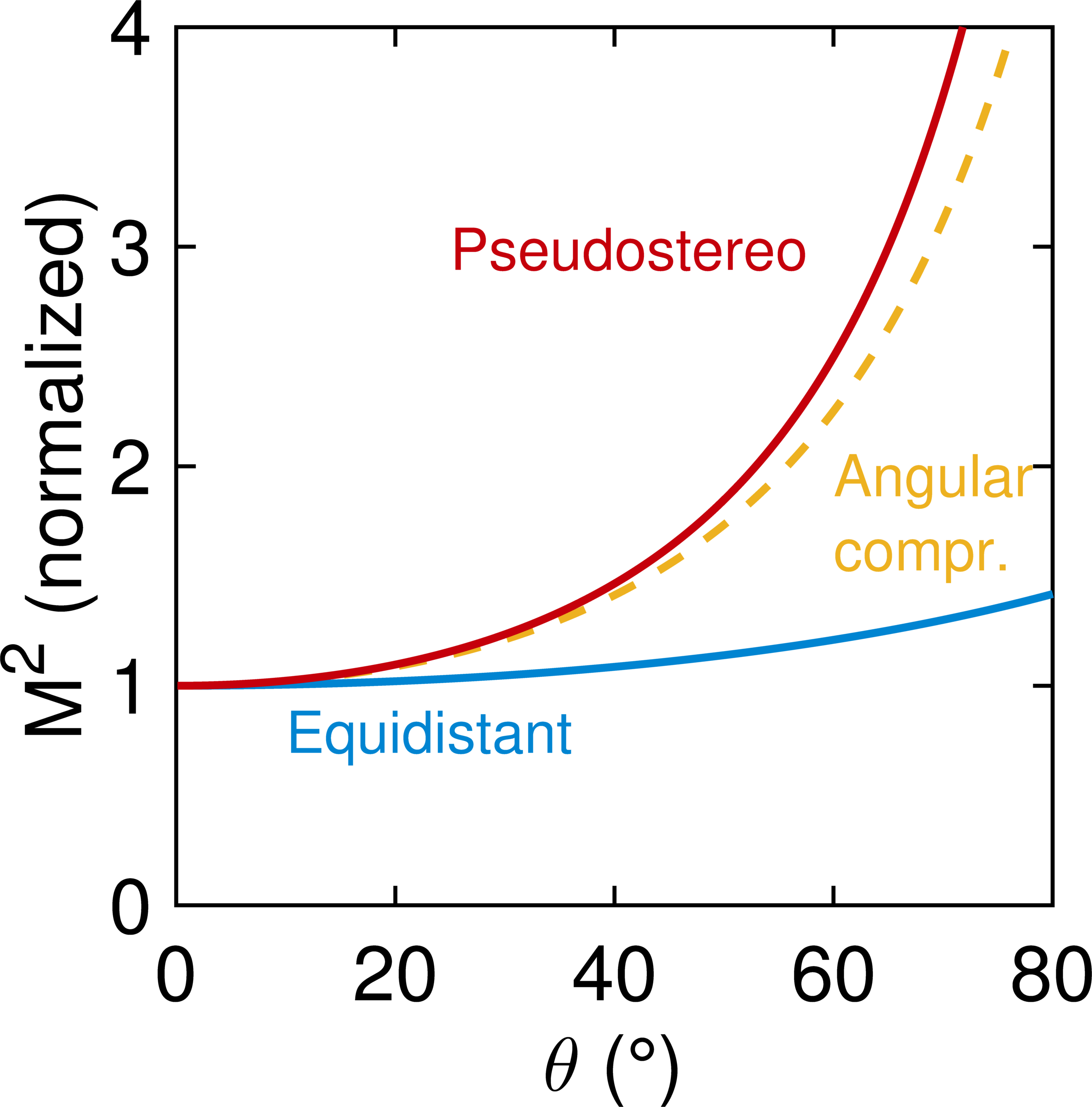}\label{comp_M}}
	\caption{ Comparison of the pseudo-stereographic and azimuthal equidistant projection models (as well as the angular compression model). (a) Distance vs. angle. (b) Normalized local image compression factor.(c) Normalized local magnification factor.}
\end{figure}

Figure~\ref{comp_Dist_vs_angle} compares the standard pseudo-stereographic projection model with the azimuthal equidistant model. For convergence at low angle, we chose $k=L/\xi$. {\revised This choice imposes that the models share the same image compression factor a low angles.} The angular compression model is also shown, and it is confirmed that, while a pseudo-stereographic projection model is not a simple angular compression, it is still well approximated by it up to wide angles. The pseudo-stereographic model and the equidistant model do diverge at high angles.

\subsubsection{Image compression factor}
In FIM and APT, the projection properties, and particularly the image compression factor, are not a constant of a specific microscope but are affected by the geometry of the specimen itself \citep{Gipson1980, hyde1994,gault_advances_2009, vurpillot2011b, Loi2013}. While it can change during a single APT analysis, it may actually not be constant throughout a single FIM image or an instantaneous desorption map. The image compression factor is defined as the ratio between the real angle $\theta$ and the apparent angle $\theta'=\mathrm{atan}\left(\frac{\rho}{L}\right)$. To measure its variation, we can define a {\revised\emph{local}} image compression factor as $\frac{\mathrm{d}\theta}{\mathrm{d}\theta'}$. Figure~\ref{comp_ICF} shows $\frac{\mathrm{d}\theta}{\mathrm{d}\theta'}$ (normalized by the low angle ICF $\xi$) as a function of $\theta$ for the projection models. It shows that in an equidistant projection model, 2 close crystallographic poles are actually closer together when they are at the edge of the detector than when they are at the center. Only for a pure angular compression (and not for the standard pseudo-stereographic model) is the local ICF constant.

\subsubsection{Magnification}
{\revised A possible quantification of the magnification as a function of the position on the detector }can be derived as follow. If one imagines a small {\revised elementary} circle at the surface of the specimen at an angle $\theta$, its radius can be expressed as $R\mathrm{d}\theta$ along one direction and $R\sin\theta\mathrm{d}\psi$. The image of this circle through the projection will be an ellipse, the 2 dimensions of which are given by $\mathrm{d}\rho$ and $\rho\mathrm{d}\psi$. We define the square of the magnification $M$ as the ratio of the area of the small circle on the detector by its area at the specimen surface
\begin{equation}
M^2=M_\theta M_\psi=\frac{\mathrm{d}\rho}{R\mathrm{d}\theta}\frac{\rho\mathrm{d}\psi}{R\sin\theta\mathrm{d}\psi}=\frac{\mathrm{d}\rho}{\mathrm{d}\theta}\frac{\rho}{R^2\sin\theta}\label{grandissement}
\end{equation}

In the general case, $M$ is a function of $\theta$, i.e. at any given moment it is not constant throughout the detector. A consequence of this is that the number of hits on an area of the detector is also a function of $\theta$ and, even in the absence of evaporation artifacts, has no reason to be uniform on the detector. 

In the limits of small angles, the local magnification $M^2$ is equal to $L^2/(\xi R)^2$. We thus plot the value of $M^2$ normalized by $L^2/(\xi R)^2$ for the different projection models (Figure~\ref{comp_M}). This figure shows that the commonly used projection model predicts that a small feature at the surface of the {\revised specimen} (for instance, the terrace of a pole) would be imaged with a significantly broader area at the edge of the detector than at the center. 
{\revised This is not typically what is observed experimentally: Figure~\ref{Desorp} is a desorption map from a pure aluminum specimen acquired on a LEAP 5000 XS instrument. 41 crystallographic poles were manually positioned and indexed. The relative size of the $(1 1 3)$ and $(1 -1 3)$ are similar, although if the true projection was close to a pseudo-stereographic projection, differences in size up to about $30\%$ would be expected for poles located at $\approx 35^\circ$ from each other. } The equidistant projection model does not predict such a large variation in surface magnification throughout the detection area. This difference in the predicted surface magnification will be of importance because of its influence on the depth of analysis model through the surface of analysis.

\section{Experimental and modeling assessment of the projection models}
\begin{figure}[htbp]
	\centering\includegraphics[width=10cm]{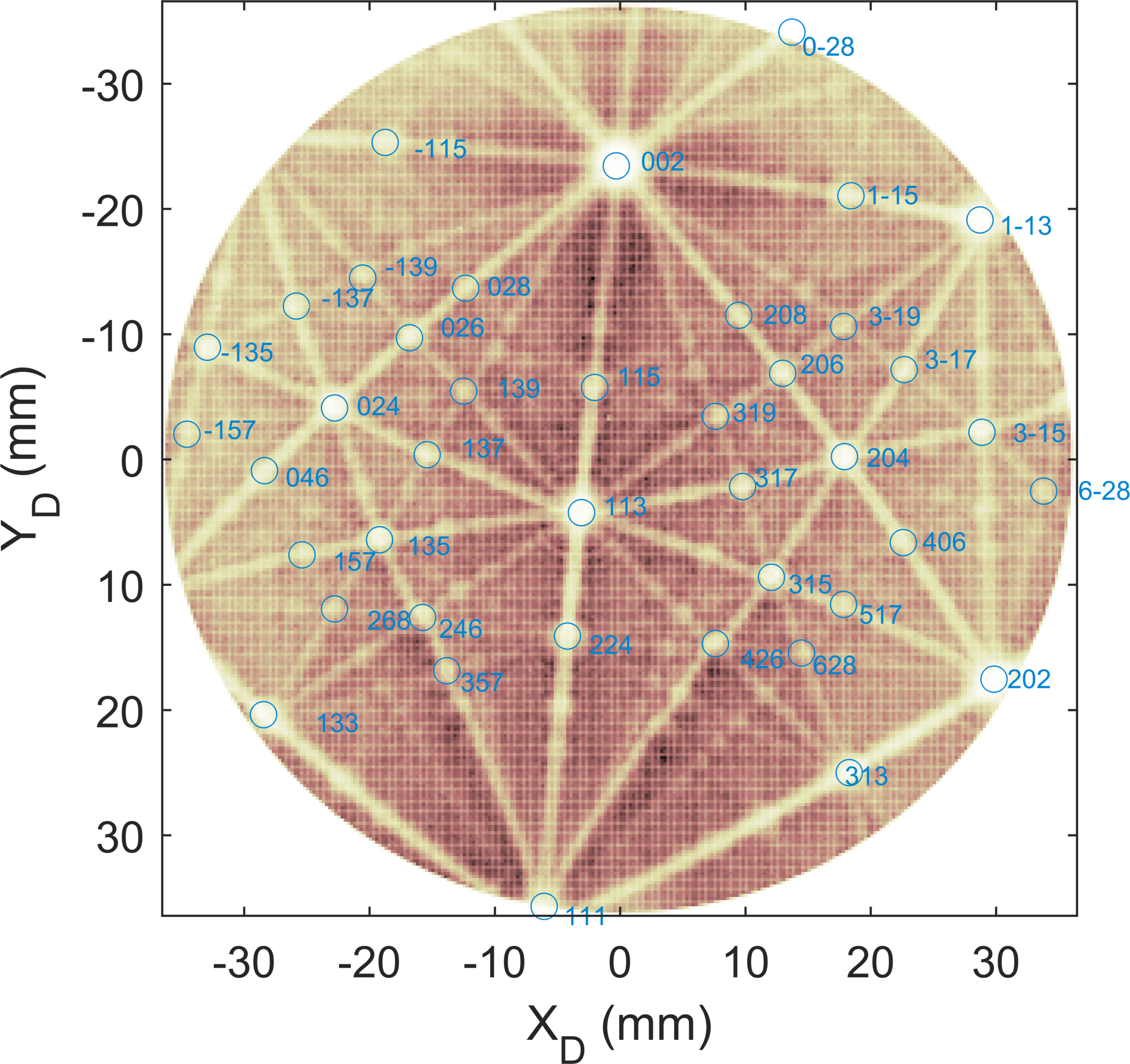}
	\caption{\label{Desorp} Desorption map of a pure aluminum sample obtained on a LEAP 5000 XS instrument with 41 indexed crystallographic poles.}
\end{figure}

An experimental assessment of the accuracy of the angular projection can be made by using either FIM images or desorption maps from an APT dataset, if it bears enough crystallographic information such as poles or zone axis. 
This is easiest with pure metals. {\revised In the set of crystallographic poles shown in Figure~\ref{Desorp}, with their position on the detector, it is possible to directly compare the observed positions to the positions predicted by the projection models.} 

\subsection{Distance vs. angle}
\begin{figure}[htbp]
	\centering
		\subfigure{\includegraphics[height=6.5cm]{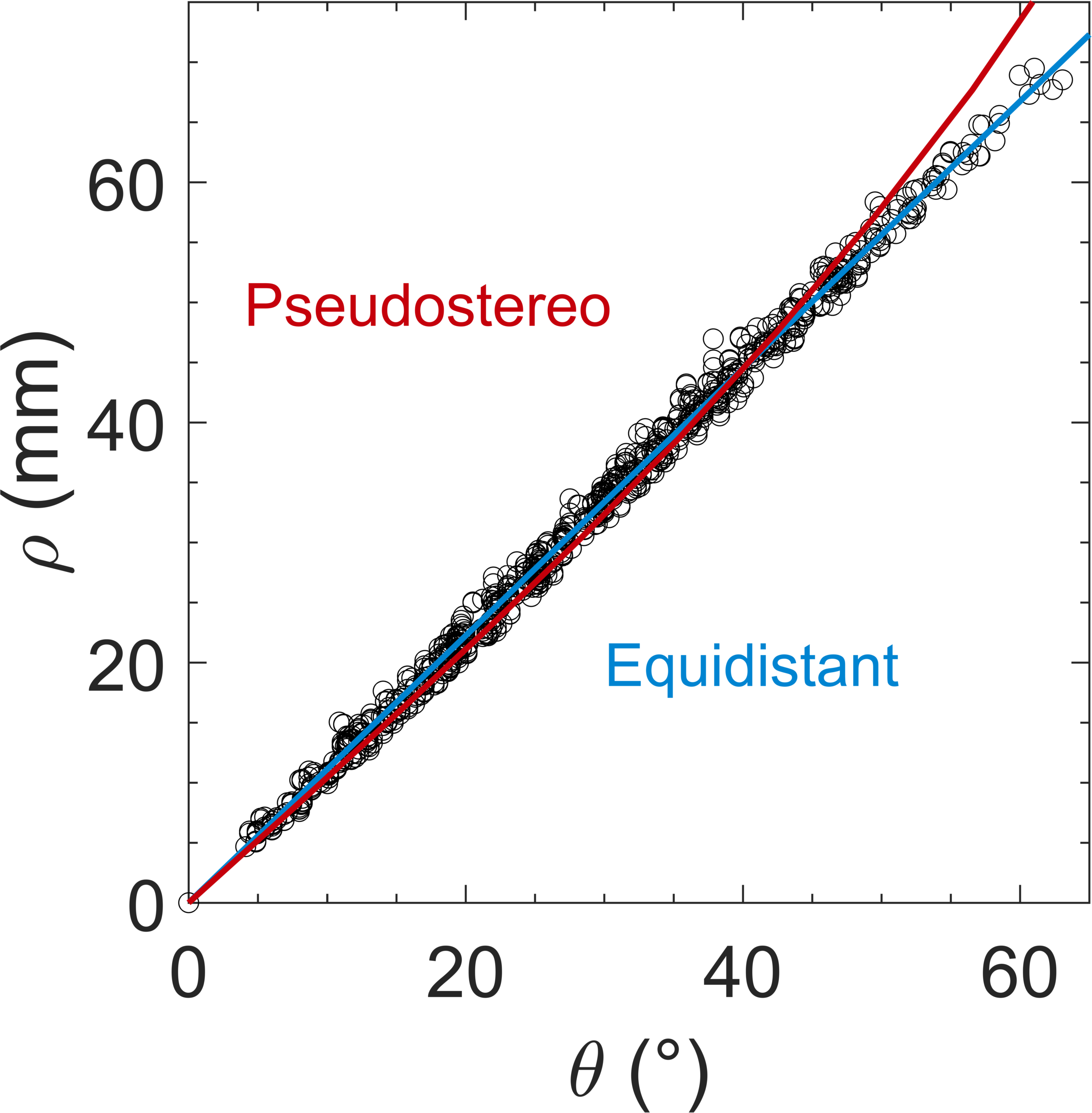}}
		
		\subfigure{\includegraphics[height=6.5cm]{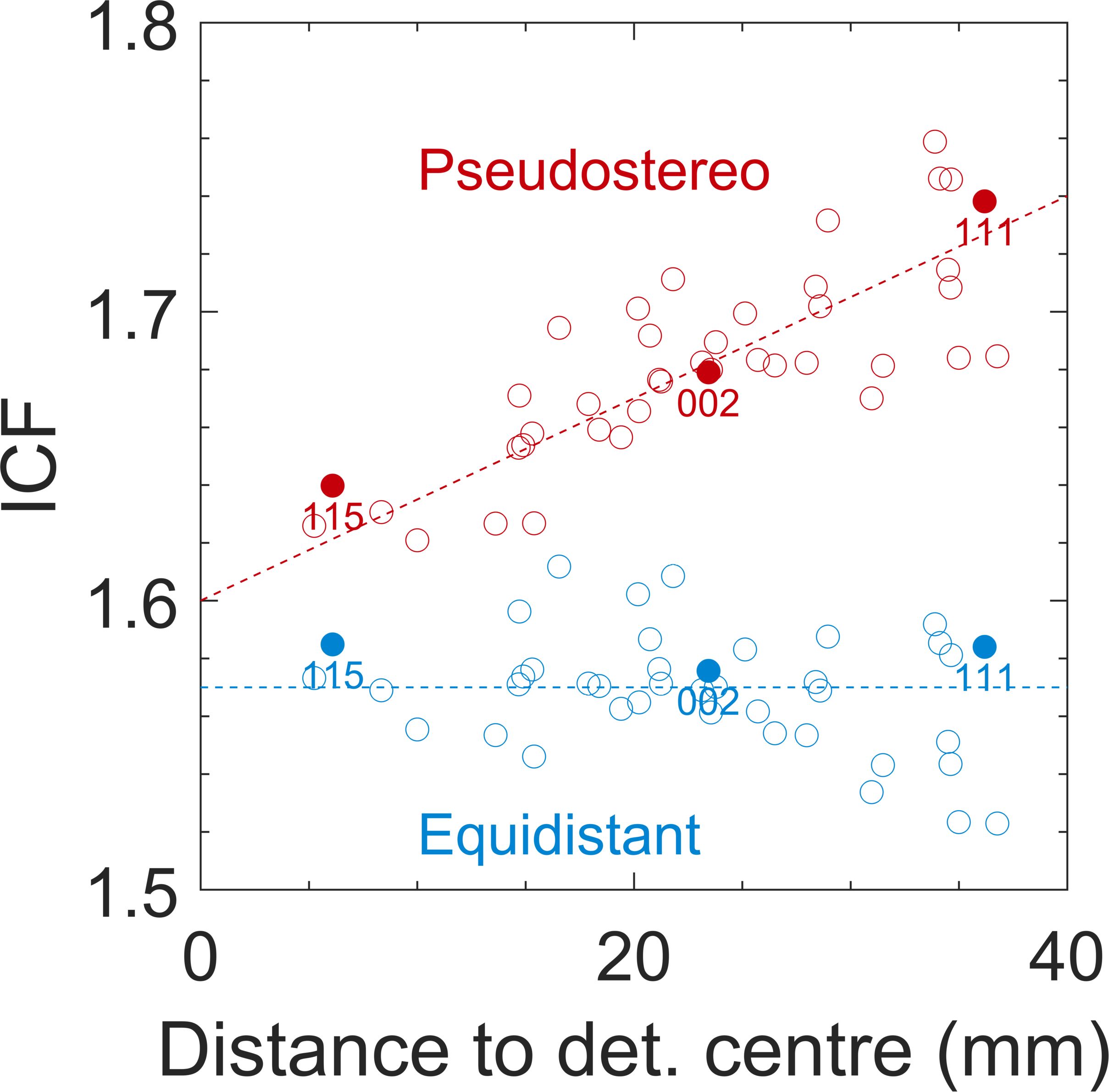}}
		\hfill		
		\subfigure{\includegraphics[height=6.5cm]{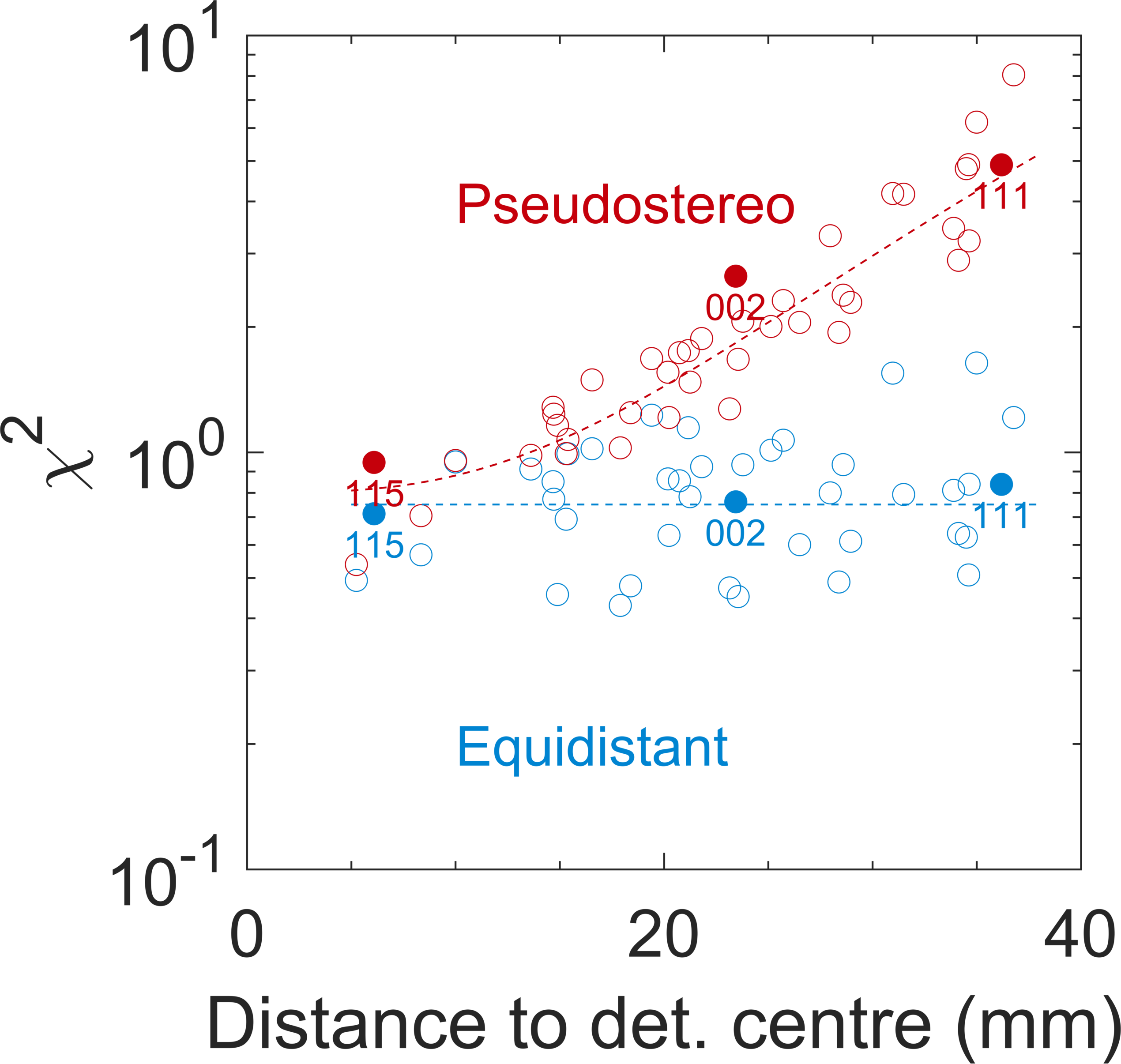}}
	\caption{{\revised (a) Distance between the centre of a set of 41 crystallographic poles observed in the analysis of a pure-Al dataset obtained on a LEAP 5000 XS (each pole considered as the projection centre). (b) Best fit ICF ($\xi=L/k$) obtained for pseudo-stereographic (red) and equidistant (blue) models respectively, as a function of the position of the pole considered as projection centre. (c) Quality of the fits obtained in (b) as assessed by $\chi^2=\frac{1}{N}\sum (\rho_\text{exp}-\rho_\text{model})^2$}}\label{rho_vs_theta}
\end{figure}

Figure \ref{rho_vs_theta}a shows the relationship between the distance on the detector and the angle between crystallographic features. To plot this figure, we have separately considered each identified pole as the center of projection, and plotted the distance between the center and all the other poles as a function of the crystallographic angle. {\revised Overlaid are two best fit curves corresponding, in blue, to an equidistant projection,  and, in red, to a pseudo-stereographic projection. Because each model has been optimized separately, the corresponding image compression factors are different (as shown, for instance, by the different slope at origin).  It is readily visible that the experimental data is linear up to very large angles and that the best fit pseudo-stereographic projection can not accurately reproduce the experimental distribution, which is particularly striking at large angles. }

{\revised Figure \ref{rho_vs_theta}b shows the image compression factors ($\xi=L/k$) resulting in the best fits for each pole considered as the projection center, as a function of the distance of that pole to the detector center. It can be seen that 1) the pseudo-stereographic model always predict a higher ICF, which arises in order to compensate for the non-linearity; and 2) the pseudo-stereographic is very dependent on the chosen projection center, in contrast with the equidistant model.  Figure \ref{rho_vs_theta}c assesses the quality of the fits by plotting $\chi^2=\frac{1}{N}\sum (\rho_\text{exp}-\rho_\text{model})^2$. It shows that the pseudo-stereographic model gives consistently less accurate results than the equidistant model. }

Such an analysis shows that (i) the azimuthal equidistant {\revised projection} closely matches the experimental data,  better than the {\revised pseudo-stereographic projection}; and (ii) {\revised the equidistant model is essentially immune (within accessible angles) to errors on the position of the projection center which is virtually always unknown.}

\begin{figure}[htbp]
	\centering
	\includegraphics[width=8cm]{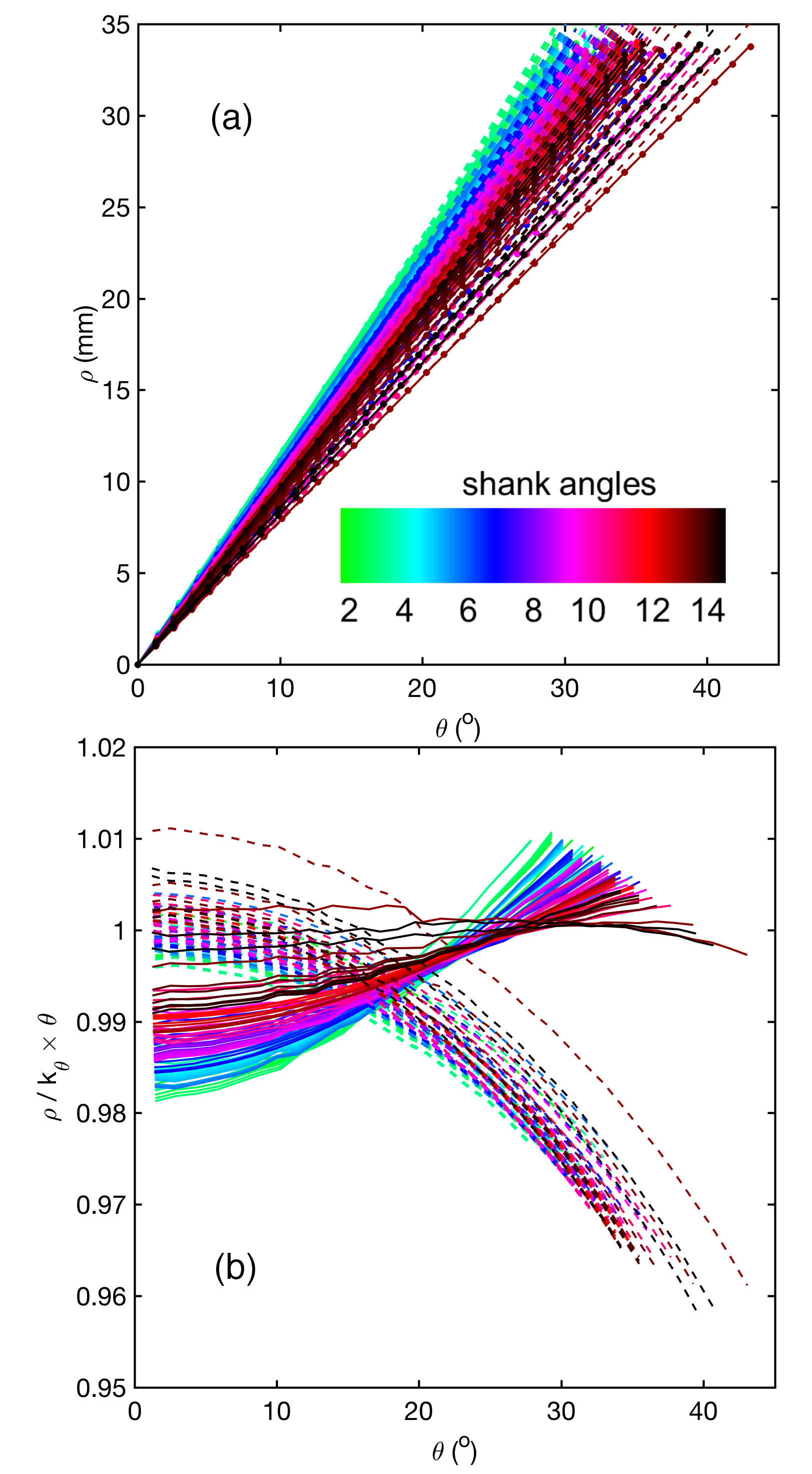}
	\caption{(a) Distribution of the distance from the ion impact position to the center of the detector $\rho$ as a function of the launch angle $\theta$ (dots) for various shank angles (see color bar) and specimen radii in the range of 20\textemdash170 nm; superimposed are a linear regression (solid line) as well as the expected distribution based on the average image compression factor (dashed line). (b) ratio of the distribution obtained from the simulations to the linear regression (solid line) as well as that based on the average image compression factor.\label{rho_vs_theta_simus}}
\end{figure}

In electrostatic simulations of ion trajectories, with a geometry mimicking a full-size commercial instrument introduced by \citet{Loi2013}, this linear relationship was also found to reproduce well the trend observed in the variation of the distance $\rho$ of an ion impact to the center of the projection, which corresponds in this case to the center of the detector, as a function of the emitting angle \citep{Larson2013a} for a variety of specimen geometries. In Figure \ref{rho_vs_theta_simus} (a), $\rho$ is plotted as a function of the launch angle for specimens with shank angles varying from 2 to 14$^{\circ}$ and radii in the range of 20 to 170 nm. {\revised For each distribution, a linear regression corresponding to an equidistant projection was calculated and is  displayed as a solid line. In addition, a dashed line is displayed which shows the distribution expected from a pseudo-stereographic projection using the estimation of the average image compression factor derived by \citet{Loi2013}. The ratio between the results of the simulations and both the linear regression and the pseudo-stereographic projection is shown in Figure \ref{rho_vs_theta_simus} (b). For the equidistant projection, this ratio only varies within a narrow range of $\pm 2 \%$ around unity with a standard deviation for the ratio $2\sigma_{equidistant}=.011$, while for the pseudo-stereographic projection, more significant variations are observed, with $2\sigma_{pseudo-stereographic}=.017$.}

\begin{figure}[htbp]
	\centering
	\includegraphics[width=6.5cm]{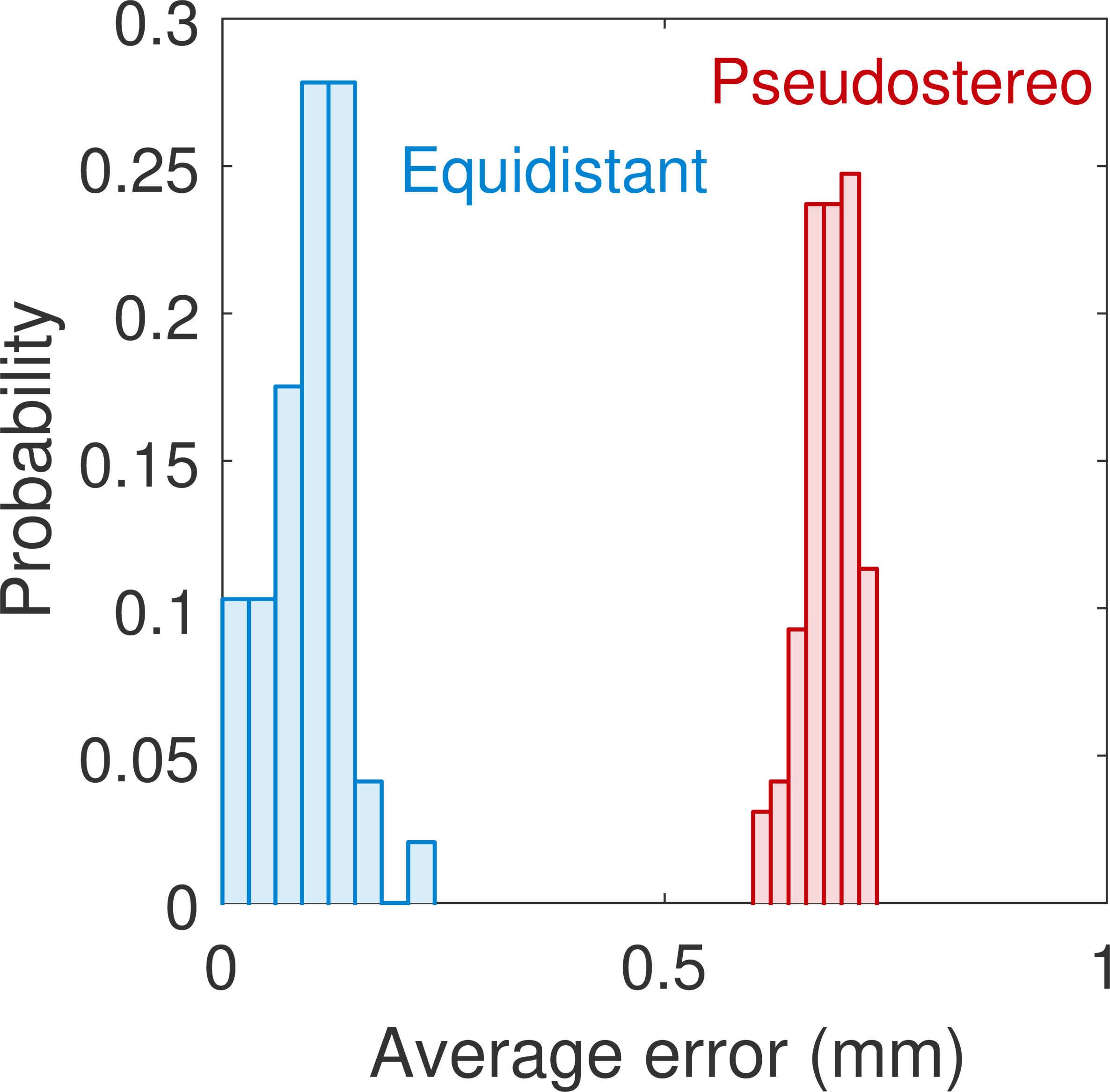}
	\caption{\label{histo_errors}Histograms of the average distance between simulated detected position and position predicted by the pseudo-stereographic (red) and equidistant (blue) models.}
\end{figure}

{\revised In order to have a better appreciation of how such variations could affect experimental data, we have calculated the average positioning errors between the simulated distance $\rho$ on the detector and the one predicted by the pseudo-stereographic and equidistant projection models for each simulation. The resulting histograms are shown on Fig.~\ref{histo_errors} for the 97 simulations. The equidistant model has a distribution of error localized around zero whereas the pseudo-stereographic projection models introduces a significant systematic error, for a wide variety of sample geometries (shank angles varying from 2 to 14$^{\circ}$ and radii in the range of 20 to 170nm).}

\subsection{Projection and orientation adjustment}
\begin{figure}[htbp]
	\centering
		\includegraphics[width=8cm]{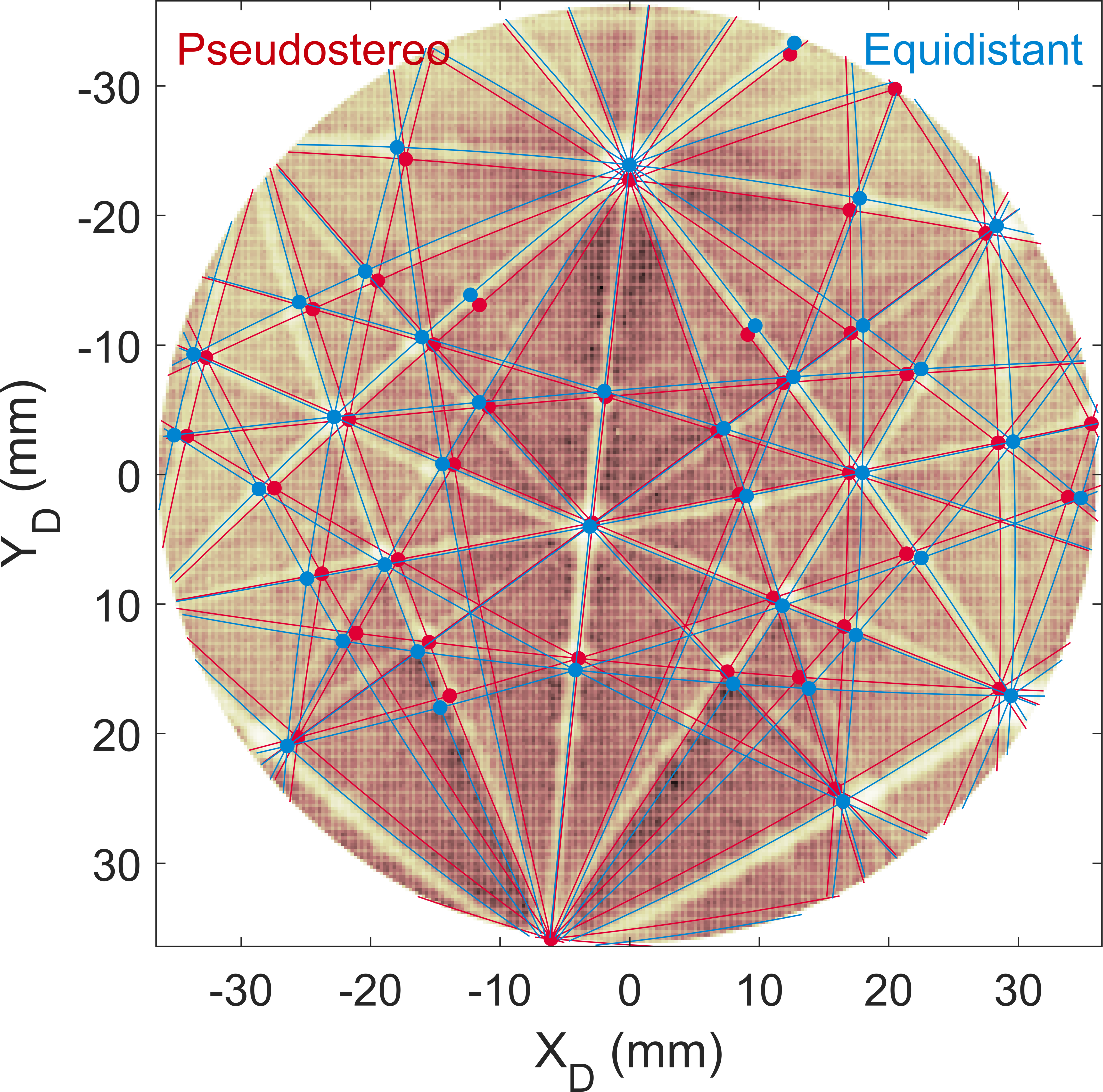}
	\caption{\label{DesorpFit} Same desorption map as in Fig.~\ref{Desorp} superimposed with the adjusted azimuthal equidistant (blue) and pseudo-stereographic (red) projections of 41 visible indexed crystallographic poles.}
\end{figure}

Since the azimuthal equidistant angular projection accurately describes the actual projection in APT despite its very simple equations, it opens the way for automatic adjustments of the crystallographic features contained in the desorption maps, such as that of Fig.\ref{Desorp}, in order to accurately obtain the orientation of the specimen. This is what was done on Fig.~\ref{DesorpFit} where we superimposed the desorption image of a pure Al specimen, shown in Fig.~\ref{Desorp}, with the predicted position of the crystallographic poles as well as zone axes based on the two projections {\revised (with ICFs corresponding to Fig. \ref{rho_vs_theta}a)}. To adjust the orientation of the specimen, we have minimized the distance between the observed and predicted poles only. An alternative methods would have been to adjust the orientation by pattern-matching on the figure showing the zone axes. For known structures, building a database that could be used for pattern-matching is feasible only if the projection is known. Such an effort would allow for automated determination of the orientation of the specimen by techniques similar to those used in electron microscopy orientation mapping, e.g. electron backscatter diffraction \citep{britton_tutorial:_2016}, nano-beam diffraction \citep{Rauch2014AutomatedTEM}, or transmission Kikuchi diffraction \citep{Keller2012,Trimby2012OrientationMicroscope}.

\section{Discussion on the effect of the projection model on the reconstructed volume}
It is difficult to determine a good metric for the accuracy of atom probe reconstructions since it really depends on what is being investigated. However, there are two general aspects that are usually, albeit implicitly, used as quality assessment of the reconstruction:
\begin{enumerate}
\item Angular distortions: if the volume contains planar features such as an interface, thin layers or platelets precipitates, one expects them to appear flat in the reconstruction. 
\item Distance accuracy: it is very common to check for known interplanar distances as a mean of reconstruction calibration.
\end{enumerate}

Both of those aspects had been discussed by \citet{bas_general_1995} already. A perfect illustration of this is the expression of $z$ in Eq.~\ref{coordSample}. It consists in the sum of $z_c$, the analyzed depth, and $R\cos\theta$, which links the curvature of the sample to the position of the atoms in the final reconstructed volume. Schematically, we can say that the error on $R\cos\theta$ is indicative of angular distortions, while the error on $z_c$ corresponds to depth calibration error. Interestingly, these two aspects are not independent. 

Errors on $R\cos\theta$ will occur if the projection model is inaccurate. Systematic errors can, to a certain extent, be compensated by an adjusted $R$ with respect to its physical value. However, this will also influence every other aspects of the reconstruction, in particular the depth of analysis \citep{gault_advances_2009}.

If we know the actual projection law, it is possible to reverse-project any given object to test the reconstruction protocol. In \ref{annexe}, we derive the analytic equations of two planes normal to the analysis direction and located at a distance $\tau$ from each other, mimicking the reconstruction of a thin layer of thickness $\tau$. It should be kept in mind that this is in the absence of any other reconstruction artifacts such as local magnification or trajectory aberrations due to variation in local evaporation fields. We can apply this to the simulated geometries used above. Again, we are in a very ideal situation where the sample has a completely smooth surface. In this context, no near-field effect can deteriorate the reconstruction \citep{Vurpillot2000a, Oberdorfer2013}, and we should thus obtain the best possible reconstructed layer. In addition, we have a full knowledge of the sample geometry, so that no calibration should be needed.

Eq.~\ref{eqPlans} shows that the expression of the reconstructed $z$ is dependent on $\theta$ (whereas it should be constant for a plane normal to the direction of analysis). Again it has 2 contributions: (i) the effect of the error on $R\cos\theta$ (proportional to $\frac{R^*}{R}\frac{\cos\theta^*}{\cos\theta}$) and (ii) the effect of the error on the depth of analysis (proportional to $\left(\frac{R}{R^*}\frac{\sin\theta_{max}}{\sin\theta^*_{max}}\right)^2$).

The calibration of an APT reconstruction is very often performed by checking a known interplanar distance, i.e. by calibrating the depth of analysis. We see from eq.~\ref{eqPlans} that for this we need to adjust $\left(\frac{R}{R^*}\frac{\sin\theta_{max}}{\sin\theta^*_{max}}\right)^2$ to make it close to 1. The influence of $\theta_{max}$ is very important here as its contribution is squared. Of course it can be adjusted by artificially adjusting $R$ and/or the image compression factor, which is what is necessarily done in practice in the absence of other information on their value, but it can only result in some degree of angular distortions. It also means that, in the case where the radius has been measured, for instance by ex-situ electron microscopy, the actual value may not be the one that should be used to optimize the calibration of the reconstruction. The same holds true for cases where the \emph{image compression factor} can be measured.

\begin{figure*}[htbp]
	\centering
	\includegraphics[width=\textwidth]{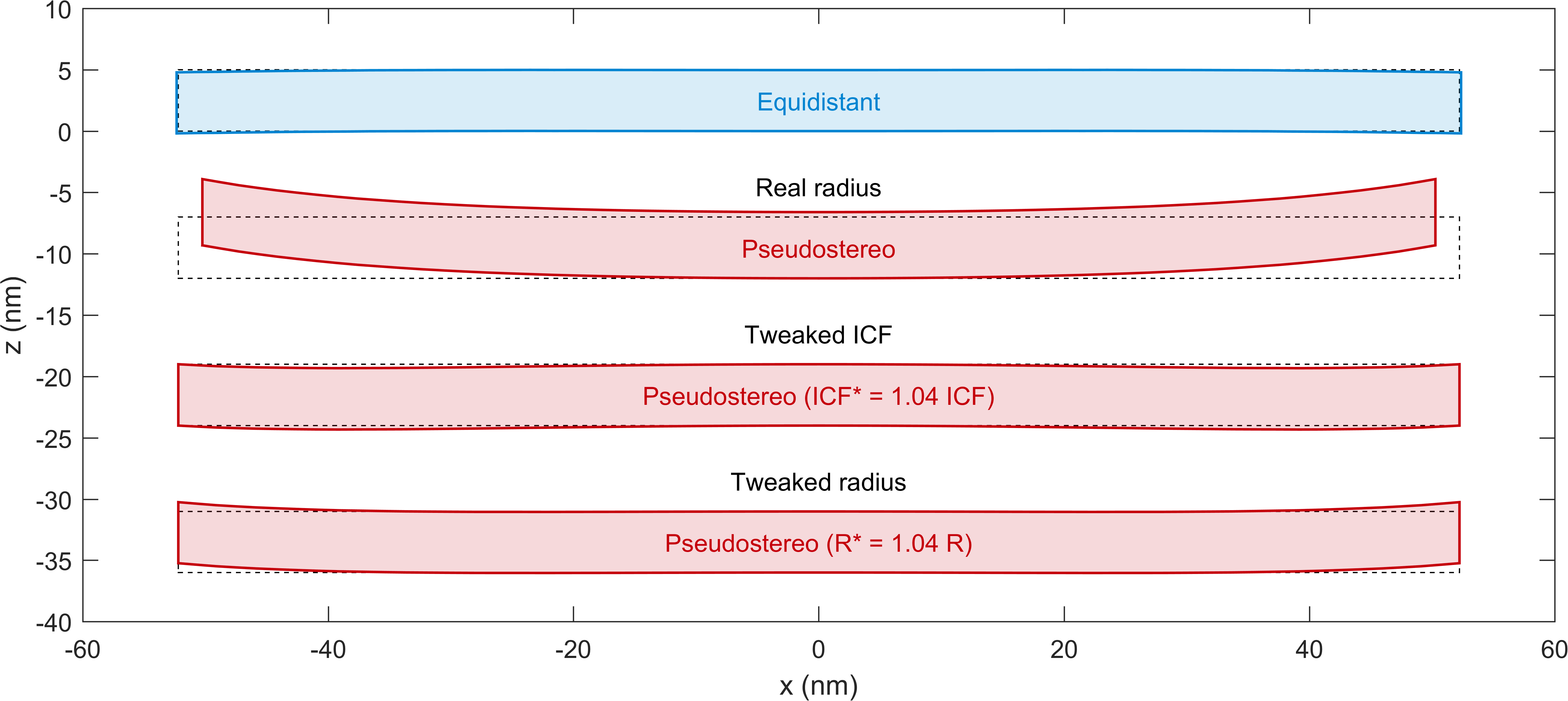}
	\caption{\label{recons_couche}Effect of the projection model on the reconstruction of a thin layer normal to the direction of analysis for a simulated geometry of $R=90$nm and $\alpha=14^\circ$. The ideally reconstructed layer is depicted as dashed black lines. The equidistant projection model (blue) gives satisfactory results, whereas pseudo-stereographic model needs adjustment of the parameters with respect to the physical parameters.}
\end{figure*}

To illustrate this, we have used eq.~\ref{eqPlans} to reconstruct a 5nm thin layer normal to the analysis in one of the simulated sample geometries ($R=90$nm, $\alpha=14^{\circ}$). We show only the trace of this layer in the $(x,z)$ plane. The result is shown on figure~\ref{recons_couche}.

With the actual physical parameters as input, the equidistant projection models quite accurately predicts the shape of the layer. Its thickness is accurate. Only faint distortions appear at large angles. The pseudo-stereographic projection predicts a layer thickness about 8\% thicker than the expected 5nm, mostly linked to a wrong value of $\theta_{max}$ that causes an inaccurate estimation of the analyzed surface. The distortions are very important, even at moderate angles. If one assumes an \emph{a priori} knowledge of the thickness of the layer, one could adjust the reconstruction parameters to achieve it. For this, one could change the value of the radius and/or that of the ICF. This is illustrated by the 2 lower images in Fig.~\ref{recons_couche} where the layer has been forced to a central thickness of 5 nm. It is obvious that there remain some distortions and the layer does not appear flat. Again, this is in a very ideal situation where the model should work at its best. This explains why, because of the chosen standard angular projection model, there can always be, even in the most accurately calibrated APT reconstruction, some remaining distortions which sadly may lead potential readers to be less confident in the results. 

\section{Conclusions}
In conclusion, we would like to point that the pseudo-stereographic projection was introduced in the early implementation of the tomographic reconstruction protocol,  and, for the small field-of-view instruments, was working fine. While it has been extended to wide angle instruments, it should in fact only be considered as a small-angle approximation of the real projections. The azimuthal equidistant model, while not ideal, has a validity domain much closer to the modern instruments setup.

In summary, we have shown that:
\begin{itemize}
\item the ion projection in atom probe tomography and field ion microscopy is best described by an azimuthal equidistant projection;
\item the azimuthal equidistant is expected to work on a broad range of specimen geometries;
\item this projection was shown to be {\revised not only more accurate, but also} more robust than the pseudo-stereographic projection when it comes to errors on e.g. the position of the center of the projection;
\item the implementation of a protocol based on such a projection is simple and could easily be generalized.

\end{itemize}

\section*{Acknowledgement}
Drs M.P. Moody, L.T. Stephenson, R.K.W. Marceau, D. Haley, T.C. Peterson, F. Vurpillot, B.P. Geiser $\&$ D.J. Larson are all thanked for fruitful discussions over the years. We extend our gratitude to Shyeh Tjing – Cleo – Loi who developed and performed the Lorentz-based simulations that have enabled part of this work. BG is grateful for the support from Profs. S.P. Ringer \& J.M. Cairney as well as the Australian Microscopy \& Microanalysis Research Facility (AMMRF) at the University of Sydney, where some of the data presented herein was obtained. Dr Eric J{\"a}gle and Mr Liang Wu are acknowledged for the provision of the pure-Al dataset obtained on the Cameca LEAP 5000. For the record, the authors owe Dr J{\"a}gle an alcoholic beverage to thank him for letting us use the dataset. Finally, BG is grateful for the support provided by Prof. Raabe and the group at MPIE. 

\section{References}


\begin{thebibliography}{20}
\expandafter\ifx\csname natexlab\endcsname\relax\def\natexlab#1{#1}\fi
\expandafter\ifx\csname url\endcsname\relax
  \def\url#1{\texttt{#1}}\fi
\expandafter\ifx\csname urlprefix\endcsname\relax\def\urlprefix{URL }\fi

\bibitem[{Bas et~al.(1995)Bas, Bostel, Deconihout, and
  Blavette}]{bas_general_1995}
Bas, P., Bostel, A., Deconihout, B., Blavette, D., 1995. A general protocol for
  the reconstruction of 3d atom probe data. Appl. Surf. Sci. 87/88, 298--304.

\bibitem[{Blavette et~al.(1982)Blavette, Sarrau, Bostel, and
  Gallot}]{Blavette1982}
Blavette, D., Sarrau, J.~M., Bostel, A., Gallot, J., 1982. {Direction and depth
  of atom probe analysis}. Revue De Physique Appliquee 17~(7), 435--440.

\bibitem[{Brandon(1964)}]{Brandon1964c}
Brandon, D.~G., 1964. {Accurate Determination Of Crystal Orientation From Field
  Ion Micrographs}. Journal of Scientific Instruments 41~(6), 373--375.

\bibitem[{Britton et~al.(2016)Britton, Jiang, Guo, Vilalta-Clemente, Wallis,
  Hansen, Winkelmann, and Wilkinson}]{britton_tutorial:_2016}
Britton, T.~B., Jiang, J., Guo, Y., Vilalta-Clemente, A., Wallis, D., Hansen,
  L.~N., Winkelmann, A., Wilkinson, A.~J., 2016. Tutorial: {Crystal}
  orientations and {EBSD} — {Or} which way is up? Materials Characterization
  117, 113--126.

\bibitem[{Cerezo et~al.(1999)Cerezo, Warren, and Smith}]{Cerezo1999a}
Cerezo, A., Warren, P., Smith, G., 9 1999. {Some aspects of image projection in
  the field-ion microscope}. Ultramicroscopy 79~(1-4), 251--257.

\bibitem[{Gault et~al.(2011)Gault, Haley, de~Geuser, Moody, Marquis, Larson,
  and Geiser}]{gault_advances_2011}
Gault, B., Haley, D., de~Geuser, F., Moody, M.~P., Marquis, E.~A., Larson,
  D.~J., Geiser, B.~P., 2011. {Advances in the reconstruction of atom probe
  tomography data}. Ultramicroscopy 111~(6), 448--457.

\bibitem[{Gault et~al.(2009)Gault, Moody, De~Geuser, Tsafnat, La~Fontaine,
  Stephenson, Haley, and Ringer}]{gault_advances_2009}
Gault, B., Moody, M.~P., De~Geuser, F., Tsafnat, G., La~Fontaine, A.,
  Stephenson, L.~T., Haley, D., Ringer, S.~P., 2009. Advances in the
  calibration of atom probe tomographic reconstruction. Journal of Applied
  Physics 105, 034913.

\bibitem[{Geiser et~al.(2009)Geiser, Larson, Oltman, Gerstl, Reinhard, Kelly,
  and Prosa}]{Geiser2009a}
Geiser, B.~P., Larson, D.~J., Oltman, E., Gerstl, S.~S., Reinhard, D.~A.,
  Kelly, T.~F., Prosa, T.~J., 2009. {Wide-Field-of-View Atom Probe
  Reconstruction}. Microscopy and Microanalysis 15 (suppl 2), 292--293.

\bibitem[{Gipson(1980)}]{Gipson1980}
Gipson, G.~S., 1980. {An improved empirical-formula for the electric-field near
  the surface of field emitters}. Journal of Applied Physics 51~(7),
  3884--3889.

\bibitem[{Hyde et~al.(1994)Hyde, Cerezo, Setna, Warren, and Smith}]{hyde1994}
Hyde, J.~M., Cerezo, A., Setna, R.~P., Warren, P.~J., Smith, G. D.~W., 1994.
  {Lateral and depth scale calibration of the position sensitive atom probe}.
  Applied Surface Science 76/77, 382--391.

\bibitem[{Keller et~al.(2012) Keller and Geiss}]{Keller2012}
Keller, R.R., Geiss, R.H. 
{Transmission EBSD from 10 nm domains in a scanning electron microscope}. Journal of Microscopy 245, 245-–251. 


\bibitem[{Larson et~al.(2013)Larson, Gault, Geiser, De~Geuser, and
  Vurpillot}]{Larson2013a}
Larson, D., Gault, B., Geiser, B., De~Geuser, F., Vurpillot, F.,  2013. {Atom
  probe tomography spatial reconstruction: Status and directions}. Current
  Opinion in Solid State and Materials Science 17~(5), 236--247.

\bibitem[{Loi et~al.(2013)Loi, Gault, Ringer, Larson, and Geiser}]{Loi2013}
Loi, S.~T., Gault, B., Ringer, S.~P., Larson, D.~J., Geiser, B.~P., 2013.
  {Electrostatic simulations of a local electrode atom probe: the dependence of
  tomographic reconstruction parameters on specimen and microscope geometry.}
  Ultramicroscopy 132, 107--13.


\bibitem[{Miller and Forbes(2014) Miller and Forbes}] {Miller2014}
(2014) Michael K. and Forbes, Richard G.
{Atom-Probe Tomography}. Springer US, Boston, MA. 978-1-4899-7429-7.

\bibitem[{Newman et~al.(1967)Newman, Sanwald, and Hren}]{Newman1967}
Newman, R.~W., Sanwald, R.~C., Hren, J.~J., 1967. {A method for indexing field
  ion micrographs}. Journal of Scientific Instruments 44, 828--830.

\bibitem[{Oberdorfer et~al.(2013)Oberdorfer, Eich, and
  Schmitz}]{Oberdorfer2013}
Oberdorfer, C., Eich, S.~M., Schmitz, G., 5 2013. {A full-scale simulation
  approach for atom probe tomography.} Ultramicroscopy 128, 55--67.

\bibitem[{Rauch and V{\'{e}}ron(2014)}]{Rauch2014AutomatedTEM}
Rauch, E., V{\'{e}}ron, M., 2014. {Automated crystal orientation and phase
  mapping in TEM}. Materials Characterization 98, 1--9.

\bibitem[{Snyder(2007)}]{Snyder1997}
Snyder, J., 2007. Flattening the Earth: Two Thousand Years of Map Projections.
  University of Chicago Press.
  
\bibitem[Stender et~al. (2007)]{Stender2007}
Stender, P.,  Oberdorfer, C., Artmeier, M., Pelka, P., Spaleck, F., Schmitz, G, 2007. 
{New tomographic atom probe at University of Muenster, Germany}. Ultramicroscopy 107(9), 726-733
    
\bibitem[{Trimby(2012)}]{Trimby2012OrientationMicroscope}
Trimby, P.~W., 9 2012. {Orientation mapping of nanostructured materials using
  transmission Kikuchi diffraction in the scanning electron microscope}.
  Ultramicroscopy 120, 16--24.

\bibitem[{Vurpillot et~al.(2000)Vurpillot, Bostel, Cadel, and
  Blavette}]{Vurpillot2000a}
Vurpillot, F., Bostel, A., Cadel, E., Blavette, D., 2000. {The spatial
  resolution of 3D atom probe in the investigation of single-phase materials}.
  Ultramicroscopy 84~(3-4), 213--224.

\bibitem[{Vurpillot et~al.(2011)Vurpillot, Gruber, Da~Costa, Martin, Renaud,
  and Bostel}]{vurpillot2011b}
Vurpillot, F., Gruber, M., Da~Costa, G., Martin, I., Renaud, L., Bostel, A., 2011. {Pragmatic reconstruction methods in atom probe tomography.}
  Ultramicroscopy 111~(8), 1286--94.

\bibitem[{Wilkes et~al.(1974)Wilkes, Smith, and Smith}]{Wilkes1974}
Wilkes, T.~J., Smith, G. D.~W., Smith, D.~A., 1974. {On the quantitative
  analysis of Field Ion Micrographs}. Metallography 7, 403--430.

\end{thebibliography}


\appendix
\section{Equation of a reconstructed plane}

\label{annexe}
For simplicity, we give only the 2D trace of a reconstructed layer in the $(x,z)$ plane such as the normal to the layer is along the axis of the tip. We will describe the plane by finding the equations of its 2 interfaces (top and bottom). Conventionally, we can assume that interface 1 is at depth $0$ and interface 2 at depth $\tau$.  In this case, the equation of the correctly reconstructed interfacial planes are at $x_1=R_1\sin\theta, z_1=0$ and $x_2=R_2\sin\theta, z_2=\tau$. 

The atoms originating from plane 1 at angle $\theta$ will correspond to a radius $R_1$ and and a depth $z_{c1}$ such as (eq.~\ref{coordSample}):
\begin{equation}
	z_1=z_{c1}+R_1\cos\theta
\end{equation}
Since $z_1=0$, we have
\begin{equation}
	z_{c1}=-R_1\cos\theta
\end{equation}

If the radius evolves with a shank angle $\alpha$ so that $R_1=R_0+w(z_{c1}-z_{c0})$, with $w=\frac{\sin\alpha}{1-\sin\alpha}$, then 
\begin{equation}
	z_{c1}=-(R_0+w(z_{c1}-z_{c0}))\cos\theta
\end{equation}
or 
\begin{equation}
	z_{c1}=\frac{w\cos\theta z_{c0}-R_0\cos\theta}{1+w\cos\theta}
\end{equation}

$z_{c0}$ is the initial depth of analysis, when the radius was equal to $R_0$. The earlier atoms from plane 1 to be detected will be those situated at $\theta=\theta_{max}$, so that $z_c(\theta_{max})=z_{c0}$. We find:
\begin{equation}
	z_{c0}=-R_0\cos\theta_{max}
\end{equation}
and finally:
\begin{equation}
	z_{c1}=-R_0\cos\theta\frac{1+w\cos\theta_{max}}{1+w\cos\theta}
\end{equation}
and similarly for plane 2:
\begin{equation}
	z_{c2}=\frac{\tau-R_0\cos\theta(1+w\cos\theta_{max})}{1+w\cos\theta}
\end{equation}

This expression gives the depth that is probed when evaporating the planes. When we reconstruct the planes, however, we use parameters which are deduced from the reconstruction model and which might be erroneous. Let us denote them $R^*$, $\theta^*$ and $\theta_{max}^*$. The reconstructed plane will be at:
\begin{equation}
	z=R^*\cos\theta^*+z_c^*
\end{equation}

$z_c^*$ can be calculated by noting that 
\begin{equation}
	\mathrm{d}z_c^*=\frac{\mathrm{d}z_c^*}{\mathrm{d}N}\frac{\mathrm{d}N}{\mathrm{d}z_c}\mathrm{d}z_c
\end{equation}
or
\begin{equation}
	\mathrm{d}z_c^*=\left(\frac{R\sin\theta_{max}}{R^*\sin\theta^*_{max}}\right)^2\mathrm{d}z_c\label{dz_star}
\end{equation}

Because of the evolution of the radius, this equation is best integrated numerically, however, to better visualize the effect of badly chosen reconstruction parameters, let us treat the conservative case of constant $R$ (i.e. $w=0$). Then eq.~\ref{dz_star} is trivially integrated and 
\begin{equation}
	z_c^*=\left(\frac{R\sin\theta_{max}}{R^*\sin\theta^*_{max}}\right)^2z_c
\end{equation}
and
\begin{equation}
	z^*=\left(\frac{R\sin\theta_{max}}{R^*\sin\theta^*_{max}}\right)^2z_c+R^*\cos\theta^*
\end{equation}

We then have the equations of the 2 planes (with $w=0$):
\begin{equation}
\begin{aligned}
	z_1^*&=R\cos\theta\left(\frac{R^*}{R}\frac{\cos\theta^*}{\cos\theta}-\left(\frac{R}{R^*}\frac{\sin\theta_{max}}{\sin\theta^*_{max}}\right)^2\right)\\
	z_2^*&=R\cos\theta\left(\frac{R^*}{R}\frac{\cos\theta^*}{\cos\theta}-\left(1-\frac{\tau}{R\cos\theta}\right)\left(\frac{R}{R^*}\frac{\sin\theta_{max}}{\sin\theta^*_{max}}\right)^2\right)
	\label{eqPlans}
\end{aligned}
\end{equation}

\end{document}